\documentclass{article}

\usepackage{Support_Docs/PRIMEarxiv}

\usepackage{amsmath,amsfonts,bm}

\usepackage{xcolor}

\def\eqref#1{equation~\ref{#1}}

\def\1{\bm{1}}

\def\vw{{\bm{w}}}
\def\vx{{\bm{x}}}
\def\vy{{\bm{y}}}

\def\mI{{\bm{I}}}
\def\mJ{{\bm{J}}}

\def\mSigma{{\bm{\Sigma}}}

\DeclareMathAlphabet{\mathsfit}{\encodingdefault}{\sfdefault}{m}{sl}
\SetMathAlphabet{\mathsfit}{bold}{\encodingdefault}{\sfdefault}{bx}{n}

\usepackage[utf8]{inputenc} 
\usepackage[T1]{fontenc}    
\usepackage{hyperref}       
\usepackage{url}            
\usepackage{booktabs}       
\usepackage{amsfonts}       
\usepackage{nicefrac}       
\usepackage{microtype}      
\usepackage{lipsum}
\usepackage{fancyhdr}       
\usepackage{graphicx}       
\usepackage{cleveref} 
\usepackage[backend=biber,natbib=true,style=authoryear-comp,]{biblatex}
\usepackage{dsfont}
\usepackage{wrapfig}

\title{Mapping Connectomic Structure to Function(s) in Cerebellar-like Networks using Kernel Regression}

\author{
  William Dorrell, Peter Latham \\
  Gatsby Computational Neuroscience Unit \\
  UCL, London \\
  \texttt{dorrellwec@gmail.com}
}

\begin{document}
\maketitle

\begin{abstract}
Cerebellar-like networks, in which input activity patterns are separated by projection to a much higher-dimensional space before classification, are a recurring neurobiological motif, present in the cerebellum, dentate gyrus, insect olfactory system, and electrosensory system of the electric fish. Their relatively well-understood design presents a promising test-case for probing principles of biological learning. The circuits' expansive projections have long been modelled as random, enabling effective general purpose pattern separation. However, electron-microscopy studies have discovered interesting hints of structure in both the fly mushroom body and mouse cerebellum. Recent numerical work suggested that this non-random connectivity enables the circuit to prioritise learning of some, presumably natural, tasks over others. Here, rather than numerical results, we present a robust mathematical link between the observed connectivity patterns and the cerebellar circuit's learning ability. In particular, we extend a simplified kernel regression model of the system and use recent machine learning theory results to relate connectivity to learning. 
We find that the reported structure in the projection weights shapes the network's inductive bias in intuitive ways: functions are easier to learn if they depend on inputs that are oversampled, or on collections of neurons that tend to connect to the same hidden layer neurons. Our approach is analytically tractable and pleasingly simple, and we hope it continues to serve as a model for understanding the functional implications of other processing motifs in cerebellar-like networks.
\end{abstract}

\section{Introduction}

\begin{figure}[h]
\includegraphics[width=\textwidth]{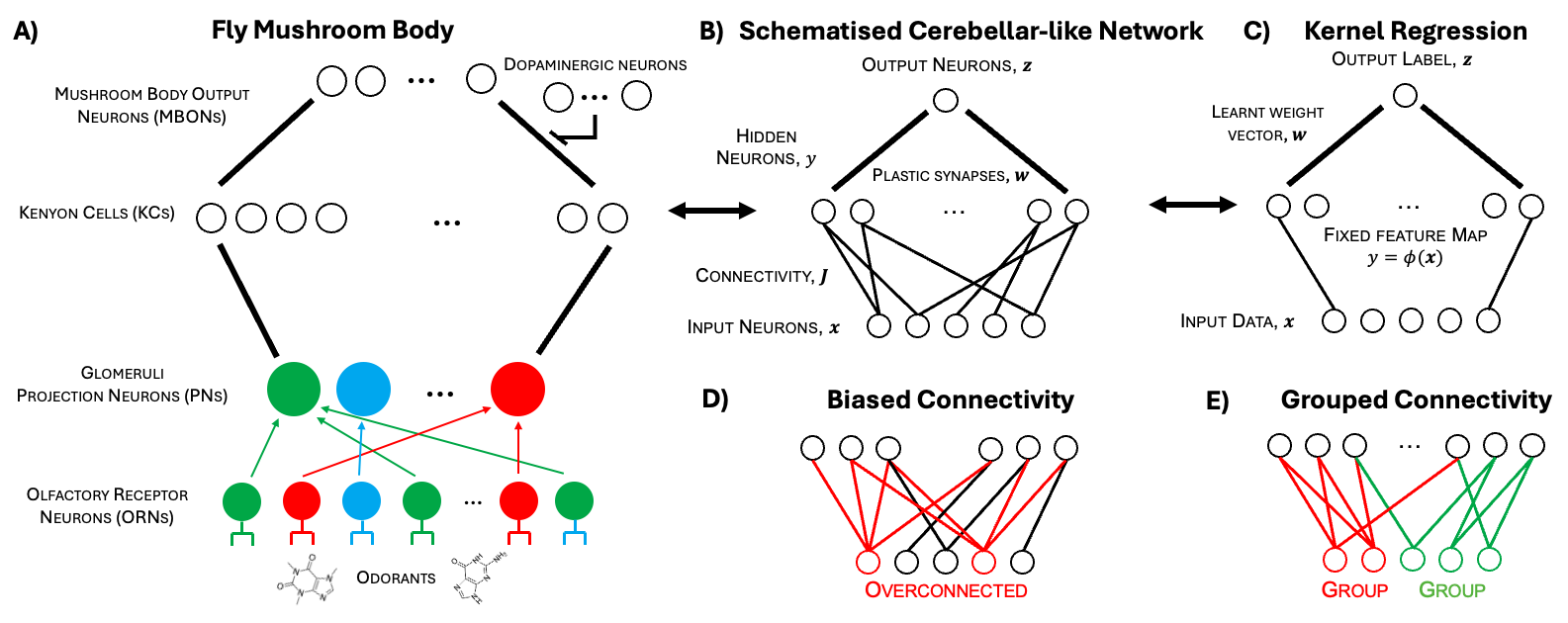}
\caption{{\bf A) Schematic of fly mushroom body circuit.} Odorants trigger activity in olfactory receptor neurons (ORNs). ORNs contain a unique receptor protein, signalled by their colour; neurons with the same receptor protein send projections to a shared glormulus. There they synapse onto projection neurons, which carry the activity to the mushroom body. In the mushroom body the dimensionality of the representation is expanded by a factor of 50 in the kenyon cells (KCs), before being assigned a valence by mushroom body output neurons (MBONs). Dopaminergic neurons encode an error signal that modulates the KC-to-MBON connectivity to ensure correct classification. {\bf B) Simplified Network} We use a model focusing only on the expansive projection and subsequent labelling which {\bf (C)} has uncanny similarities with the kernel regression algorithm. We consider two structural motifs - \textbf{(D)} biased connectivity, in which some input neurons connect more than others; and \textbf{(E)} grouped connectivity, in which the inputs are grouped: if a hidden layer neuron is connected to one member of the group it is likely to be connected to others.}
\label{fig:cerebellum-network_schematics}
\end{figure}

The crystalline nature of cerebellar cortex has attracted theoretical work for over 50 years, since it was first posited to perform pattern separation \citep{marr1969theory,albus1971theory,ito1972neural, kawato202150}. The basic processing involves a feed-forward passage of information from a small number of input mossy fibres, through an expansion to a vastly larger population of granule cells which activate sparsely, and themselves converge onto a small number of output purkinjie cells \citep{eccles2013cerebellum,shadmehr2020population}, \cref{fig:cerebellum-network_schematics}B. Versions of this expand-sparsify-contract circuit motif have since been recognised in the dentate gyrus \citep{borzello2023assessments}, the mushroom body of the fly \citep{modi2020drosophila}, and the electrosensory organ of the electric fish \citep{bell1981efference}. The classic Marr-Albus-Ito model has understood these circuits via pattern separation: through the expansive and sparse mapping to the granule cell layer, potentially overlapping inputs patterns are separated, allowing supervised learning on the granule-to-purkinjie cell weights to produce the correct output for each input. To first order, this pattern separation machine is a good model; indeed, genetic perturbations that enrich or degrade the separability of inputs in the fly mushroom body directly improve or impair flies' odour discrimination \citep{ahmed2023hacking}.

However, recent data have driven updates to this theory. For example, the finding that sometimes granule cell representations are not so sparse \citep{badura2017cerebellar,knogler2017sensorimotor} conflicts with existing pattern separation models, leading to theories that relate the sparsity level to the smoothness of the learnt input-output mapping \citep{xie2022task}. 
A similar ongoing update comes from advances in the availability of connectomic data, information about which neurons are connected in both the fly mushroom body \citep{zheng2022structured} and a section of mouse cerebellum \citep{nguyen2023structured}. Previous work assumed random connectivity between input and expansion layer \citep{marr1969theory,litwin2017optimal}, a choice that produces high-dimensional representations, perfect for pattern separation. The connectomic data, however, contains signs of weak structuring. Rather than connecting randomly to input neurons, expansion layer neurons connect preferentially to some inputs~\cref{fig:cerebellum-network_schematics}D, and the inputs appear to be grouped such that if an expansion neuron is connected to one input it is very likely to be connected to other inputs in the same group,~\cref{fig:cerebellum-network_schematics}E. In the mushroom body this connectivity structure arises independently of neural activity \citep{hayashi2022mushroom}, suggesting it is genetically hard-wired. We are then prompted to ask, why is the nervous system investing effort to establish this connectivity? What functional role does it play in the pattern separating circuit?

Two recent modelling works have tackled exactly this question. \citet{zavitz2021connectivity} built models of the fly mushroom body with and without the observed structured connectivity motifs. Then they trained and tested each network on a battery of tasks, and found that the networks with the observed connectivity performed better at some tasks, and worse at others. In particular, over-connecting to some inputs allowed easy identification of single odours that activated that input, while input groupings made the network better at classifying odours that activated all input neurons within a group, tasks the fly presumably has to perform regularly. In previous work we instead  meta-learned the input-output maps that the circuit found easy to learn with or without the observed connectomic patterns \citep{dorrell2023meta}. From this we derived similar conclusions without the need to pre-specify a battery of tasks for testing the network, instead allowing a search from the full space of tasks for those the network performs best at.

In this work we present an alternative analytic route to understand the functional effect of structured connectivity that avoids the need for extensive simulations. We make use of an established correspondence between cerebellar-like circuits and an algorithm called kernel regression \citep{harris2019additive,xie2022task}. 
Recent advances in machine learning theory have characterised the generalisation properties of kernel regression; i.e. how well the algorithm is able to capture input-output mappings from limited training examples \citep{sollich1998learning,bordelon2020spectrum,simon2021neural}. We will apply these theoretical advances to a cerebellum-like model to understand the effect of different connectivity patterns on the generalisation properties of the network. In particular, using this link we assign a normative role to the structured connectivity through how it allows the circuit to learn some, presumably important, classifications quicker than if the connectivity had been random.

We begin in~\cref{sec:cerebellum_circuit_backgroun} by describing the structure of cerebellar-like circuits and how they relate to kernel regression. In~\cref{sec:cerebellum_intuition} we didactically outline theoretical results on kernel regression, and in~\cref{sec:cerebellum_analytic} we use these to analytically understand the impact of connectivity on the network's behaviour. Finally, in~\cref{sec:cerebellum_biological} we show that these core results generalise to less schematised models.

\section{Cerebellar-like Networks as Kernel Regression}\label{sec:cerebellum_circuit_backgroun}

We begin by outlining our simplified cerebellar-like model and its relationship to kernel regression.

\paragraph{Cerebellar-Like Circuits} As the main target for both connectomic work and our investigations, we will describe the fly mushroom body,~\cref{fig:cerebellum-network_schematics}A, before later relating it to other cerebellar-like networks. Odours arrive at the periphery of the fly olfactory system and activate a panel of olfactory receptor neurons (ORNs) through interactions with their receptor proteins \citep{modi2020drosophila}. There are many different types of receptor proteins, but each ORN has only one (the different coloured neurons in \cref{fig:cerebellum-network_schematics}A). The axons of ORNs containing each type of receptor converge onto a single glomerulus, of which there are about 50 in total. In the glomerulus, the olfactory receptor neurons synapse onto projection neurons (PNs) that carry the information onward, including to the mushroom body. In the mushroom body they pass information onto the $\sim$2000 kenyon cells (KCs), which form a large sparse representation of the odours that will serve as the basis for classification. Each KC receives input from roughly 7 projection neurons \citep{zheng2022structured} and a large inhibitory neuron synapses onto all the KCs and ensures that only 5-10\% of neurons are active at any one time \citep{lin2014sparse}. Finally, mushroom body output neurons (MBONs) connect to the entire KC population and are thought to represent an odour classification, signalling various dimensions of the valence. Most of the synapses in this circuit are not thought to change on short timescales, barring the important site of learning, the KC-MBON connections \citep{barnstedt2016memory}. Changes in these synapses are governed by dopaminergic neurons, whose activities are thought to encode errors in recent classifications \citep{aso2014neuronal}. Hence, this system forms a neat odour classification device, schematised in~\cref{fig:cerebellum-network_schematics}A.

This design is an example of the broader class of cerebellar-like circuits. The core features of such circuits are the expansive nonlinear projection, which separates and pre-processes the inputs through sparsification, and the subsequent downstream classification via plastic synapses modulated by an explicit error signal. Unsurprisingly, this structure is cleanly exhibited by the cerebellum itself, where mossy fibres, granule cells, and purkinjie cells play the roles of input, expansive, and output layers - with an estimated 30 times more granule cells than mossy fibres. Each purkinjie cell has an associated climbing fibre which carries an error signal: if the climbing fibre activates, it triggers a complex spike in the purkinjie cell which leads to plasticity at the granule-to-purkinjie synapses, teaching the system to correctly classify inputs. Somewhat similar patterns are seen in the electrosensory lobe of the mormyrid fish \citep{kennedy2014temporal}, and the dentate gyrus \citep{borzello2023assessments}, though in all cases there remain significant puzzles we overlook \citep{cayco2019re}, and these circuits display plasticity at other synapses \citep{d2009timing,bliss1973long}, which we ignore.

\paragraph{Simplified Model} We are interested in how structure in the expansive projection affects a cerebellar circuit's ability to learn. We therefore use a simplified model that targets this question directly. All assumptions are made for analytic ease, and some will be relaxed in later sections that show our conclusions generalise to a more realistic model. 

Our simple model contains only an expansive projection and nonlinearity, followed by linear classification. We keep the kenyon cell nonlinearity to match the observed sparsification, but drop output nonlinearities for simplicity. We assume that each input activity pattern, corresponding to the activity of glomeruli in the mushroom body or mossy fibres in the cerebellum, uses the same amount of neural activity, which we implement by assuming they are sampled uniformly from the surface of a sphere -- effectively assuming a normalisation and whitening pre-processing step \citep{carandini2012normalization,wanner2020whitening}. We'll tend to use only three input dimensions for visualisation, but our analytic results generalise to higher dimensions. This input activity is then projected to a much larger population through a connectivity matrix, $\mJ$, each element of which encodes the connection strength between a pair of input and hidden layer neurons. An elementwise nonlinearity, $\phi$, is applied to produce the final expansive layer activity,
\begin{align}
\vy = \phi(\mJ \vx) \, . 
\end{align}
We will get some analytic insight using a simple ReLU nonlinearity, but will show the same conclusions hold using a more biologically plausible sparsification so that only 5-10\% of neurons are active at one time, as observed in the fly mushroom body \citep{lin2014sparse}. Finally, a readout weight, $\vw$, assigns a label to each pattern,
\begin{align}
\hat{z} = \vw \cdot \vy \, .
\end{align}
The readout weight is trained on a set of `true' input-label pairs: $\mathcal{D} = \{\vx_i, z_i\}$, gathered during the animal's recent experience. We will choose $\vw$ using ridge regression, i.e., we choose $\vw$ to minimise a combination of the reconstruction error and size:
\begin{equation}
    \mathcal{L}(\vw) = \mathbb{E}[(\hat{z} - z_i)^2] + \lambda||\vw||_2^2 = \mathbb{E}[(\vw \cdot \phi(\mJ \vx_i) - z_i)^2] + \lambda||\vw||_2^2
\end{equation}
Choosing $\vw$ this way enables tractability while approximating the training of $\vw$ via online dopamine-driven updates.

This model matches directly onto a machine learning algorithm called kernel regression, which also comprises a nonlinear fixed processing step, followed by a linear classification using ridge regression, \cref{fig:cerebellum-network_schematics}C. This correspondence will let us harness theoretical work on kernel regression to understand the impact of connectivity choices on learning.

\section{Connectivity's impact on Learning - Intuition from Kernel Regression}\label{sec:cerebellum_intuition}

This work seeks to provide an intuitive yet rigorous exploration of the impact of connectivity structure on learning in cerebellar-like networks. In this section we introduce the relevant machine learning theory ideas for making this link.

\begin{figure}[!ht]
    \centering
    \includegraphics[width=0.8\linewidth]{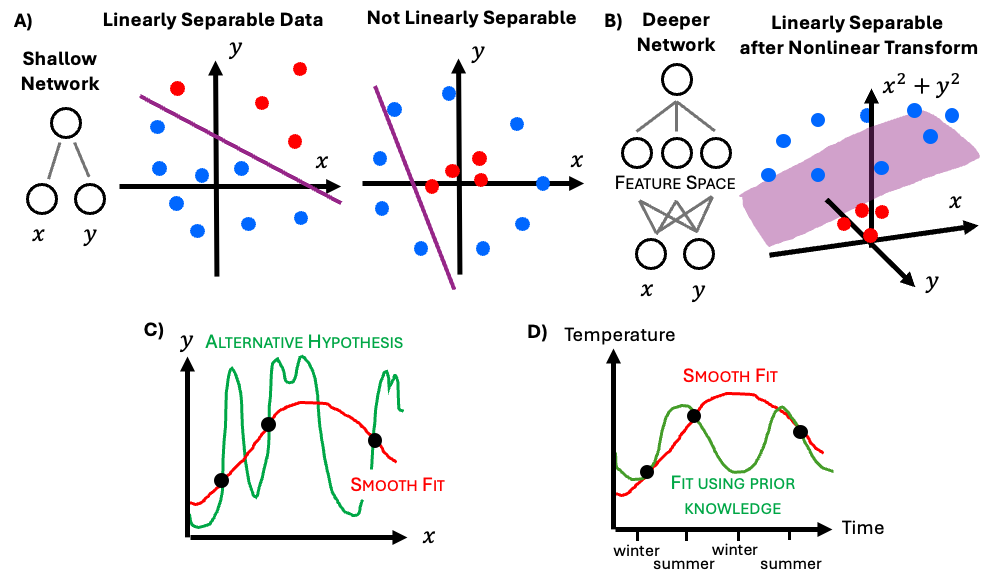}
    \caption{{\bf A)} A single linear layer can only classify linearly seperable data. {\bf B)} However adding a fixed first layer of nonlinear processing can permit a linear readout layer to perform nonlinear classifications. {\bf C)} Given any finite dataset there are infinitely many possible generalisations to unseen data. Without prior assumptions there is no reason to choose between them, i.e. from training data alone both curves are equally reasonable. {\bf D)} Generic prior assumptions might prefer a smooth solution, while prior knowledge on how temperature varies throughout the year would select a solution with a period of 12 months.} 
    \label{fig:cerebellum-kernel_schematics}
\end{figure}

\paragraph{Algorithmic Background} Kernel regression is a supervised learning algorithm; like all such algorithms, it uses a set of examples to learn an input-to-output mapping. Perhaps the simplest widely used supervised learning system is linear regression, where the predicted output, $\hat{z}\in\mathbb{R}$, is a linear function of the input, $\vy\in\mathbb{R}^N$: $\hat{z} = \vw \cdot \vy$. The weights, $\vw\in\mathbb{R}^N$, are learnt from training data in a way that hopefully generalises to unseen `test' data. This can be well understood theoretically, and forms the bedrock of all regression analyses; however it suffers from a clear shortcoming: not all outputs are linearly related to their inputs, ~\cref{fig:cerebellum-kernel_schematics}A. How can we model biology's ability to learn nonlinear relationships between inputs and outputs? One now-pervasive answer is artificial neural networks, which, through iterated layers of nonlinearities, learn sophisticated mappings. Unfortunately, these networks are not well understood, nor is it clear how learning in such deep architectures could be performed biologically \citep{lillicrap2020backpropagation}.

Kernel regression, sitting between these two extremes, avoids the above flaws. As discussed, kernel regression involves first projecting the input data through a fixed nonlinear mapping, called a feature map, then linearly classifying the features (rather than the inputs directly). By choosing an appropriate feature map, nonlinear classifications can be performed,~\cref{fig:cerebellum-kernel_schematics}B; indeed, for many choices of feature maps every possible input-output function can be learnt \citep{micchelli2006universal}. Yet kernel regression balances this flexibility with both analytic tractability and biological plausibility -- and provides an uncanny match to cerebellar-like systems. Learning only happens at the output weights and could be driven by a simple error signal -- a motif that matches findings across cerebellar-like networks where error signals often drive plasticity at expansion-to-output synapses \citep{waddell2013reinforcement,d2014organization,bell1993storage,bell1997physiology}. Further, its relative simplicity will allow us to precisely understand the impact of connectivity on learning. 

\paragraph{Inductive Bias} Having chosen our learning algorithm and established that it is capable of learning nonlinear input-output mappings, we are left with the following question: in realistic settings, with small amounts of training data, how will it perform? 

Kernel regression, like all learning algorithms, carries with it an inductive bias -- a preference for learning certain functions over others. During learning the network adapts to fit a few labelled training examples; however, having fit a finite dataset there are infinitely many ways to generalise to unseen examples,~\cref{fig:cerebellum-kernel_schematics}C, and choosing amongst them without prior assumptions is impossible. Yet, somehow, the kernel regression algorithm chooses. How it does so is its inductive bias, an embodiement of the algorithm's implicit prior assumptions about the world. A learning algorithm will only be useful if it can learn the functions it is likely to need using a reasonably small amount of data, in other words, if its prior assumptions about the world are reasonable.

A simple illustrative example is a bias towards smoothness, built into almost all learning algorithms. Consider the dataset in~\cref{fig:cerebellum-kernel_schematics}C. In reality it might have come from a high frequency function, or any of the infinitely many other functions that interpolate the training data. But, having observed only a couple of datapoints, it would seem unreasonable to guess a fast-oscillating output. Rather, without any more information, the best choice would seem to be a smooth low-frequency interpolation between the training data. If, however, you knew the input was time and the output was temperature in an unknown city, your prior assumptions would be shifted, and you might instead predict an output with a period of 12 months,~\cref{fig:cerebellum-kernel_schematics}D.

\paragraph{Kernels} The inductive bias therefore becomes a way to understand a learner's behaviour. The link between cerebellar-like networks and kernel regression is appealing because the inductive bias of kernel regression can be precisely understood using the kernel function. The {\it kernel} in kernel regression is a function that measures the similarity between two inputs. In our context it refers to the similarity between two inputs after being mapped through the nonlinear feature map of the cerebellar-like expansive projection:
\begin{equation}
    k(\vx, \vx') = \phi(\mJ \vx)\cdot\phi(\mJ \vx')
\end{equation}
In neuroscience language this corresponds exactly to the representational similarity of two inputs in the expansion layer. This is the key quantity both generally for kernel algorithms, and for our discussion. Just as different cerebellar-networks are specified by their choice of nonlinear projection, different kernel regression algorithms are defined by their choice of kernel/representational similarity. 

In short, the punchline is as follows. Kernel regression finds it easier to learn functions that assign similar labels to similar inputs. In this context, similarity between inputs is quantified by the kernel: inputs are similar if their expansion layer representations are similar.

To repeat this vital point, if two inputs, $\vx$ and $\vx'$, are represented similarly in the feature map (meaning $k(\vx,\vx')$ is large) then the kernel regression algorithm finds it easier to learn a function which assigns them similar labels. We can understand this by imagining a network trained to correctly classify a single training example $\{\vx_1, z_1\}$. Kernel regression learns the min-norm weight vector that correctly classifies this point:
\begin{equation}
    \text{Correct Classification:} \quad \vw\cdot\phi(\mJ\vx_1) = \vw\cdot\vy_1 = z_1 \qquad \rightarrow \qquad \text{Min-norm Weight Vector:} \quad \vw = \frac{z_1\vy_1}{||\vy_1||^2} \propto z_1\vy_1
\end{equation}
The resulting weight vector aligns with the datapoint's expansion-layer representation, $\vy_1$. Having learnt to correctly label a single datapoint, we can examine how it chooses to classify other unseen points:
\begin{equation}
    \hat{z}(\vx) = \vw\cdot\phi(\mJ\vx) \propto  z_1\phi(\mJ\vx_1)\cdot\phi(\mJ\vx) = z_1 k(\vx_1, \vx)
\end{equation}
This means that when presented with unseen test points the system will generalise what it has learnt in a manner that depends exactly on the similarity between the test and training point as quantified by the kernel. Similar ideas generalise to networks trained on multiple datapoints. 

In summary, any network must choose how to generalise what is has learnt from training examples. Kernel regression chooses its generalisation according to its inbuilt similarity measure, the kernel function --- $k(\vx, \vx')$ --- which measures representational similarity. If the true input-output function, $z(\vx)$, matches the structure assumed by the kernel then this generalisation will be appropriate, and the network will require few training examples to achieve low generalisation error. If, however, the algorithm's assumptions are wrong then the network will generalise to unseen regions poorly.

\paragraph{Theoretical Clarity} Pleasingly, theoretical work has made the preceding intuition precise for the kernel ridge regression learning algorithm \citep{sollich1998learning,bordelon2020spectrum,simon2023eigenlearning}. These works identify a basis in which to decompose and understand these problems: the kernel eigenfunctions, $v_i(\vx)$. Each of these functions can be thought of as an assignment of a label, $v_i(\vx)$ to each input, $\vx$. These eigenfunctions are defined analogously to eigenvectors:
\begin{equation}\label{eq:kernel_eigenfunction_defn}
\int k(\vx, \vx') v_i(\vx') p(\vx')d\vx' = \lambda_i v_i(\vx)
\end{equation}
where $\lambda_i$ is their eigenvalue and $p(\vx)$ is the distribution of inputs. These functions form an orthonormal basis, meaning that any function learner is learning, $z(\vx)$, can be decomposed as a weighted sum of the eigenfunctions:
\begin{equation}\label{eq:eigendecomposition}
    z(\vx) = \sum_{i=1}^\infty c_i v_i(\vx)
\end{equation}
where $c_i$ are the co-ordinates of $z(\vx)$ in this space. 

Crucially, the generalisation of the learner on $z(\vx)$, the target function, can be understood by instead understanding the generalisation on the eigenfunctions $v_i(\vx)$. It turns out that the learning of each eigenfunction is controlled almost entirely by the corresponding eigenvalue. The learnability of a function, $z(x)$, is defined as the alignment between $z(x)$ and the function learnt by kernel ridge regression, $\hat z(x)$, averaged over different training datasets of a given size sampled from $z(x)$ (for details see \citet{simon2023eigenlearning}):
\begin{equation}
    \mathcal{L}(z) = \mathbb{E}[\langle z, \hat z\rangle]
\end{equation}
this quantity varies between $0$ and $1$, corresponding to the cases where the learner has learnt either nothing or everything about the classification, $z(x)$. \citet{simon2023eigenlearning} find that the learnability of an eigenfunction is:
\begin{equation}
    \mathcal{L}(v_i) = \frac{\lambda_i}{\lambda_i + \kappa}
\end{equation}
where $\kappa$ is an 'effective regularisation' constant that encapsulates regularisation and number of training points: it decreases the more training points you get. In other words, eigenfunctions with higher eigenvalue are easier to learn, meaning fewer datapoints are required to reach a given accuracy, and more data always helps. The learnability of an arbitrary function (i.e. not an eigenfunction) can be calculated using the eigendecomposition,~\cref{eq:eigendecomposition}:
\begin{equation}
    \mathcal{L}(f) = \sum_i \mathcal{L}(v_i)c_i^2
\end{equation}
Hence, this theory provides an answer to our initial question: functions that align with high-eigenvalue eigenfunctions are easy to learn.

\begin{wrapfigure}{r}{0.5\textwidth}
  \begin{center}
    \includegraphics[width=0.48\textwidth]{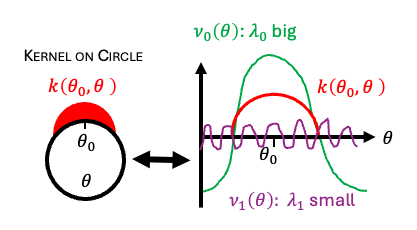}
  \end{center}
  \caption{Consider a kernel defined on a circle, points that are nearby are similar, as shown by the red curve which measured similarity between each angle $\theta$ and $\theta_0$. It is largest at $\theta_0$ and decays further away. An eigenfunction that constructively interferes with the kernel similarity, green, with have a large eigenvalue, while destructive interference, purple, produces a small eigenvalue.}
  \label{fig:kernel_eigenfunctions}
\end{wrapfigure}

As such, there is one final piece of the conceptual groundwork to be laid: the meaning of high eigenvalues. In short, an eigenfunction has high eigenvalue if it assigns similar labels to the inputs deemed similar by the kernel. This can be seen from a small sketch and the definition, a repeat of~\cref{eq:kernel_eigenfunction_defn}: 
\begin{equation}
\int k(\vx, \vx') v_i(\vx') p(\vx')d\vx' = \lambda_i v_i(\vx)
\end{equation}
This says that the value of the eigenfunction at $\vx$ is equal to a scaled version of the sum of the eigenfunction at all other points, weighted by the kernel. If, for all points where the similarity is large and positive, the labelling, $v(\vx')$, is also large and positive then they will constructively interfere, producing a large output, and correspondingly a large eigenvalue,~\cref{fig:kernel_eigenfunctions} green. On the other hand, if the kernel is large and positive across a region in which the label, $v(\vx')$, oscillates between positive and negative it will produce a correspondingly smaller output, and a small eigenvalue,~\cref{fig:kernel_eigenfunctions} purple.

This logic matches with the previous claim that functions that assign similar labels to similar points are easier to learn; for eigenfunctions, this is because they have high eigenvalue.

\paragraph{Conclusion} These results give us a way to answer our initial question: how does connectivity effect the learner's abilities? The connectivity matrix changes the structure of the representation in the expansion-layer. This can be summarised through changes to the kernel function, which measures representational similarity in the expansion-layer. These changes to the kernel cause corresponding changes to the eigenstructure. Since the learner's inductive bias can be understood through the eigenfunctions and their eigenvalues, by studying the changing eigenstructure, we can understand the inductive bias of the network as a whole. We can therefore understand the functional role of different connectivity structures by looking at their effect on the network's eigenstructure.

\section{An Analytic Link Between Connectivity and Inductive Bias}\label{sec:cerebellum_analytic}

In this section we model the observed connectivity motifs and derive their influence on the inductive bias of the network. In general, calculating a kernel's eigenstructure analytically is challenging, so for tractability we use a simple connectivity model. The last section generalises these findings to more realistic settings.

\begin{figure}[!hbt]
    \centering
    \includegraphics[width=0.65\linewidth]{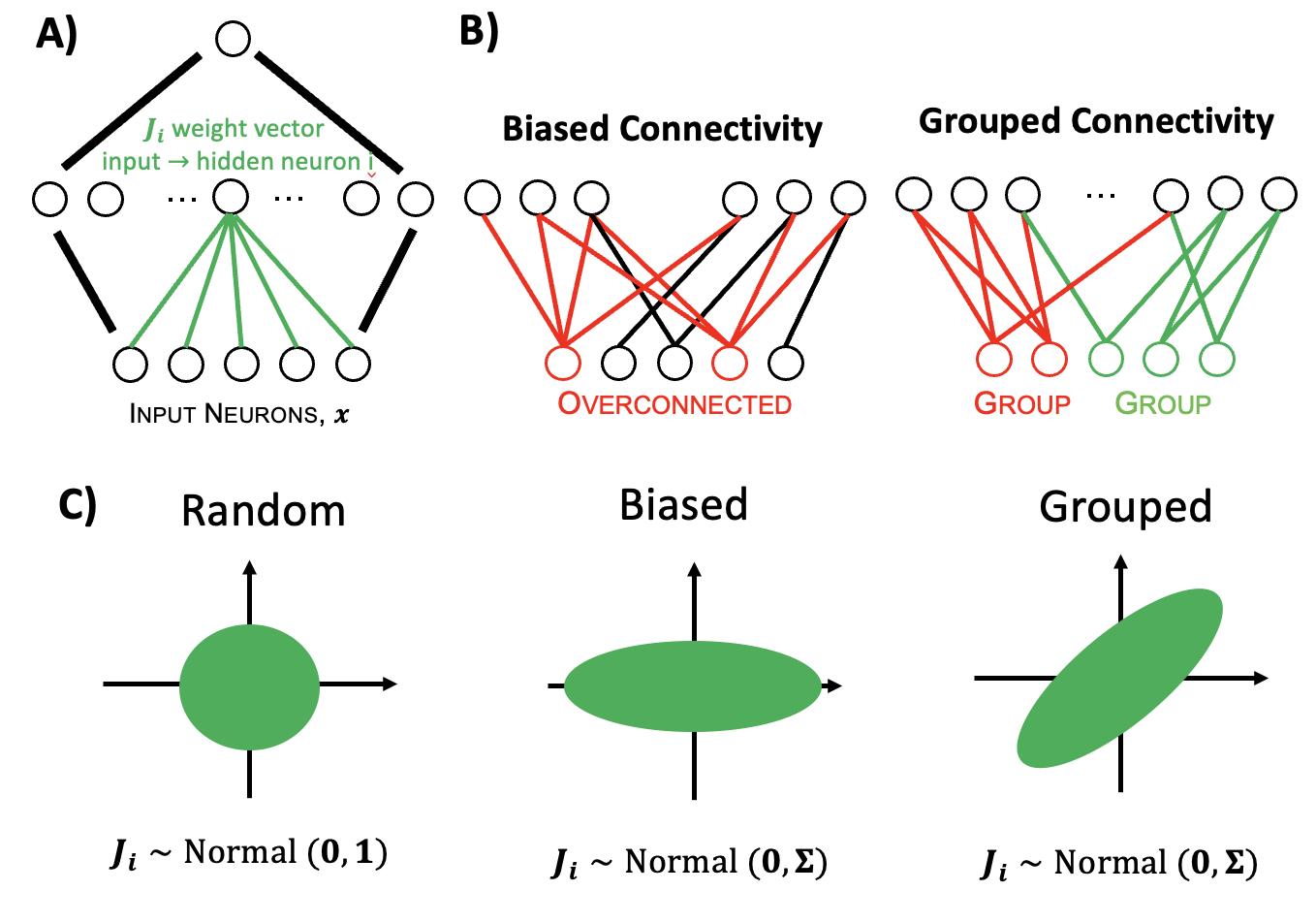}
    \caption{{\bf A)} We sample each expansion layer weight vector, $\mJ_i$, independently. {\bf B)} We model two structural motifs: 1) biased connectivity, in which some input neurons connect more than others; and 2) and grouped connectivity, in which the inputs are grouped; i.e., if a hidden layer neuron is connected to one member of the group it is likely to be connected to others. {\bf C)} We build an analytically tractable model of these structures using multivariate Gaussians. Random connectivity is modelled by sampling each weight from a Gaussian with mean zero and variance 1; biased connectivity is implemented by increasing the variance of one dimension; grouping is implemented by correlating each hidden neuron's connection to members of the input neuron grouping, meaning that if it has a strong positive connection to one member of the group, it likely will to the others.}
    \label{fig:cerebellum-connectivity_schemes}
\end{figure}

\paragraph{Connectivity Schemes} Call the vector of input weights to the $i$th expansion-layer neuron $\mJ_i$,~\cref{fig:cerebellum-connectivity_schemes}A. Our model specifies the connectivity matrix by specifying a probability distribution on $\mJ_i$, from which each expansion-layer neuron samples its input weights. We assume there are infinitely many expansion-layer neurons (one could call it `the infinitely wide nose assumption'), which is not a bad assumption in these expansive networks. To enable tractability, we constrain the distribution of $\mJ_i$ to be a multivariate Gaussian:
\begin{equation}
\mJ_i \sim \mathcal{N}(\mathbf{0}, \mSigma) \, .
\end{equation}
This minimal model allows us to study the effects of the observed connectivity structures through changing the covariance matrix, $\mSigma$. Most simply we have the null model, random connectivity, $\mJ_R$, corresponds to $\mSigma = \mI$, the identity matrix, and each weight is sampled independently from all the others. 

To study the two observed connectomic effects we introduce two simple changes to $\mSigma$. Empirically, expansion-layer cells connect to some input neurons with higher probability than others \citep{zheng2022structured,nguyen2023structured}. We model that by stretching the covariance along one input axis relative to the others,~\cref{fig:cerebellum-connectivity_schemes}C, meaning that $\mSigma$ remains diagonal, but with different values on each diagonal. This means the corresponding input neuron will have larger weights, and hence more influence on the representational similarity.

Similarly, there are correlations in the connectivity: the projection neurons form groupings such that if a kenyon cell is connected to one member of a group it is likely to be connected to the others \citep{zheng2022structured}. In our simple model with only three input neurons we'll group together one pair of them. To reflect this in connectivity we introduce correlations into $\mSigma$ such that if a kenyon cell is strongly connected to one of the paired projection neurons it is likely to also be strongly connected to the other,~\cref{fig:cerebellum-connectivity_schemes}C. This means that $\mSigma$ is blocked according to the input groups:
\begin{equation}\label{eq:correlated_sigma}
    \mSigma = \begin{bmatrix}
        1 & \rho & 0 \\ \rho & 1 & 0 \\ 0 & 0 & 1
    \end{bmatrix}
\end{equation}
\begin{wrapfigure}[15]{r}{0.5\textwidth}
  \begin{center}
    \includegraphics[width=0.48\textwidth]{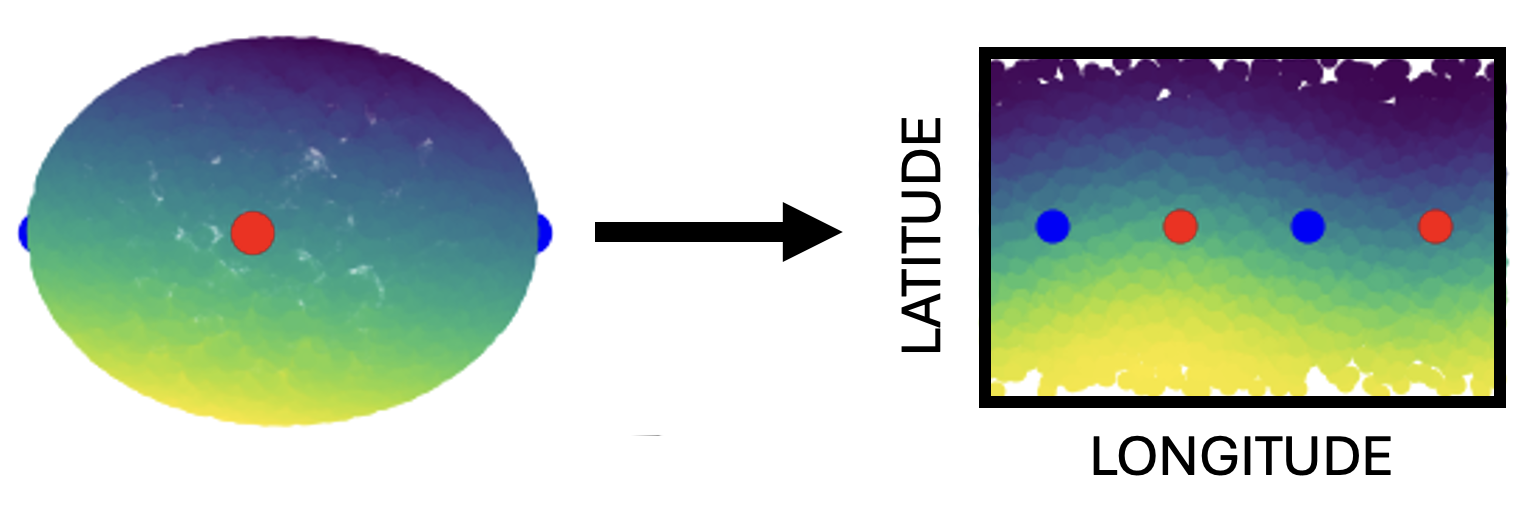}
  \end{center}
  \caption{Each eigenfunction maps from points on the sphere to a label, which we display as a colour, left. To display the full function, we project to 2D space (for map nerds, we use the equirectangular projection), and show the label as a function of latitude and longitude. For ease of orientation, the four of the points lie on the $x_1$ and $x_2$ neuron axes are shown by the red and blue points.}
  \label{fig:spherical_projection}
\end{wrapfigure}
Given all these assumptions a simple extension of the well known result of Cho \& Saul \citep{cho2009kernel} allows us to derive the kernel for each connectivity scheme, as in previous work \citep{pandey2021structured}. We will use this to understand each connectivity pattern's effect on the inductive bias.

\paragraph{Guide to Plots} With three input neurons and input activity vectors with unit norm, the input data lives on the surface of a sphere, and the input-output functions map from the sphere to a label. For visualisation, we display the eigenfunctions via a projection to 2D, showing the labelling as a colour varying with longitude and latitude,~\cref{fig:spherical_projection}. In this projection, $x_3$, the third input neuron, is mapped to the latitude, while the $(x_1,x_2)$ plane becomes the equator, as shown by the highlighted points,~\cref{fig:spherical_projection}.

\begin{figure}[!ht]
    \centering
    \includegraphics[width=\linewidth]{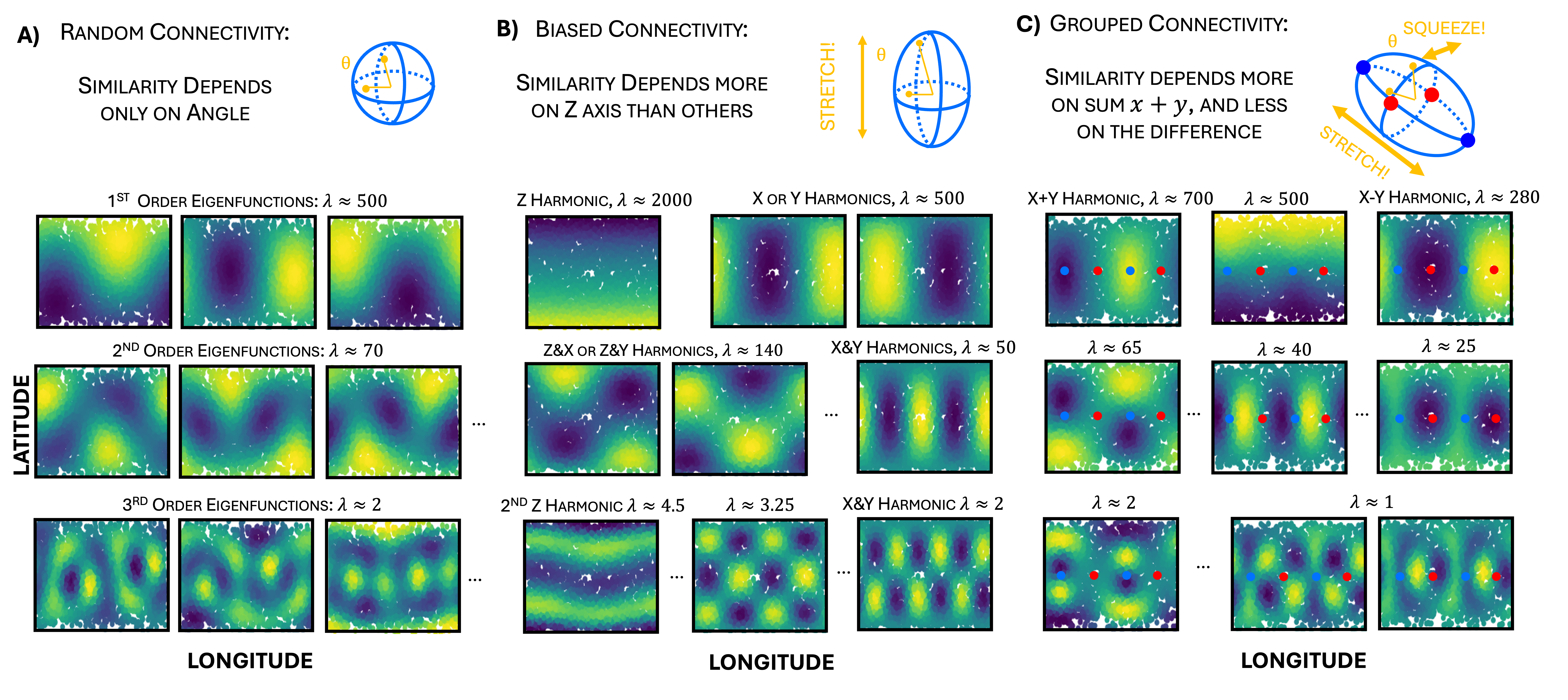}
    \caption{{\bf A)} With unbiased random connectivity the similarity depends only on the angle between points on the sphere. The eigenfunctions are assignments of a label to each point on the sphere, we therefore display them projected into 2D space,~\cref{fig:spherical_projection}. The eigenfunctions of this kernel are known to be spherical harmonics \citep{muller2012analysis}. Spherical harmonics fall into groups, there are $2l+1$ members of each group where $l$ is the order of the group, for example the first group contains 3 functions, each is a smooth oscillation pointing in one of the three 3 orthogonal directions. The subsequent orders contain more functions, more oscillations, and have a lower eigenvalue, so are harder to learn. {\bf B)} Biased connectivity can be understood by stretching the sphere along the overconnected axis -- which we've chosen to align with the latitude on the sphere, along the vertical axis. This breaks the degeneracy between spherical harmonics of the same order: functions that oscillate along the latitude-direction have higher eigenvalue and are correspondingly easier to learn than functions that oscillate in the other directions. {\bf C)} We implement grouped connectivity by positively correlating the size of the weight vector along the $x$ and $y$ directions. The corresponding kernel can be understood as measuring similarity along an ellipse stretched in the $x+y$ direction (blue dots) while being squeezed in the $x-y$ direction (red dots). This similarly breaks the degeneracy amongst spherical harmonics of the same order: variations along the $x+y$ direction (e.g. top left) have a higher eigenvalue, while variations along the $x-y$ direction have a lower one (top right). The blue points correspond to the points on the ellipse along the stretched $x+y$ direction, and the red along the squeezed $x-y$ direction.}
    \label{fig:cerebellum-eigenfunctions}
\end{figure}

\paragraph{Inductive Bias of Randomly Connected Network} For a random connectivity matrix, $\mJ_R$, the kernel depends only on the angle, $\theta$, between inputs, $\vx$ and $\vx'$ \citep{williams1998computation,cho2009kernel}:
\begin{equation}
\label{eq:kern}
    k(\vx, \vx') = \frac{|\vx||\vx'|}{\pi}[\sin(\theta) + (\pi - \theta)\cos(\theta)]
    = \frac{1}{\pi}[\sin(\theta) + (\pi - \theta)\cos(\theta)] \, .
\end{equation}
The eigenfunctions of this kernel are spherical harmonics \citep{muller2012analysis}, the extension of sine and cosine to a sphere. The higher the frequency the lower the eigenvalue and hence the harder the eigenfunction is to learn,~\cref{fig:cerebellum-eigenfunctions}A. Therefore with $\mJ_R$ the network embodies a simple smoothness prior: it generalises the labels it has learnt smoothly, and the smoother the true function is the easier it is to learn.

\paragraph{Impact of Connectivity on Kernel} The kernel under the two modified connectivity schemes,~\cref{fig:cerebellum-connectivity_schemes}B, can be written using the same formula, eq.~\ref{eq:kern}, but using a different definition of $\theta$ and $\vx$. First, the data is stretched by the square root of the covariance matrix, then~\cref{eq:kern} is applied to the resulting data.
\begin{equation}
\label{eq:kern_II}
    \tilde\vx = \Sigma^{\frac{1}{2}}\vx \qquad k(\vx, \vx') = \frac{|\tilde\vx||\tilde\vx'|}{\pi}[\sin(\tilde\theta) + (\pi - \tilde\theta)\cos(\tilde\theta)]
\end{equation}
In other words, $\tilde{\theta}$ is the angle between $\Sigma^{\frac{1}{2}} \vx$ and $\Sigma^{\frac{1}{2}} \vx'$, and $\tilde\vx$ is the length of $\Sigma^{\frac{1}{2}} \vx$. From this simple kernel we can numerically calculate the eigenfunctions. We now discuss the interpretable effects that arise from this modification to the kernel.

\paragraph{Effect of Biased Connectivity} When $\mSigma$ is diagonal, the modified kernel,~\cref{eq:kern_II}, amplifies the effect of variations along the axes with larger diagonal elements. This means that the over-connected input neuron, $x_3$ in our example, have a larger effect on the expansion-layer representational similarity. This is intuitive, overconnected neurons have more influence. This change manifests intuitively in eigenstructure: eigenfunctions with variation along the overconnected input direction are easier to learn, and hence have higher eigenvalue than those along orthogonal direction,~\cref{fig:cerebellum-eigenfunctions}B.

This leads us to our first intuitive conclusion. Networks that overconnect to some inputs find it easier to learn functions that depend upon variation of that input relative to the others. This matches the claim that monodours that perhaps only activate one projection neuron can still be well identified if they stimulate an overconnected projection neuron \citep{zavitz2021connectivity, dorrell2023meta}.

\paragraph{Effect of Correlated Connectivity} Correlations in the connectivity can be similarly understood by rotating into the eigenbasis of the relevant $\mSigma$,~\cref{eq:correlated_sigma}. Since we correlated the $x_1$ and $x_2$ connections, the highest eiganvalue eigenvector aligns with the $x_1 + x_2$ direction, and the lowest with the $x_1-x_2$ direction. Therefore, variations along $x_1+x_2$, which reflect changes in the average activity of the group, have an amplified effect on the representational similarity, while the effects of variations within the group are diminished. These changes to the kernel have a similarly intuitive effect on the eigenstructure,~\cref{fig:cerebellum-eigenfunctions}C. The network is inductively biased towards labellings that assign similar valences to datapoints that activate the entire group of projection neurons in concert,~\cref{fig:cerebellum-eigenfunctions}C. Conversely, labellings that require differentiating between inputs patterns that vary in how they distribute activity within the group while maintaining the same sum are much harder for the network to generalise.

Hence, we reach our second intuitive conclusion. Correlations in the connectivity encourage generalisation across the activity of members of the connected group, making it easy to distinguish odours that tend to activate the whole group, and hard to make distinctions between inputs that activate some members of the group but not others \citep{zavitz2021connectivity, dorrell2023meta}.

\section{More Realistic Models}\label{sec:cerebellum_biological}

The model presented above is clean, permits analytic derivation of the eigenstructure in any dimensionality. However, it suffers from some obvious mismatches with biology. We'll try and correct some of those mismatches now and demonstrate that the same broad conclusions hold. While these changes mean we cannot analytically calculate the kernel, we can use numerics to derive similar quantities. For low-dimensional inputs it is very easy to densely sample the input space, compute the kernel implied by the network on these points, and then numerically calculate the eigenfunctions. The simplicity of this approach makes it a great first pass to understand how changes to the network affect its learning ability -- to encourage adoption of this approach we share all our code, which should be easily adapted\footnote{Code at: \url{https://github.com/WilburDoz/Infinitely_Wide_Noses}}.

\paragraph{Biologically Plausible Model} In particular we change two things. First,  We remove the ReLU nonlinearity and instead, inspired by the large inhibitory Anterior Paired Lateral neuron that reciprocally inhibits the entire kenyon cell population in the fly mushroom body, modify the representation such that only the top 10\% of cells are active at a given moment \citep{lin2014sparse}.

And second, rather than using all-to-all input-to-hidden connectivity and modulating their strengths to encode structural motifs, we create a sparse connectivity matrix. In the random model we connect pairs of neurons at a fixed probability; we encode overconnection to a given input neuron by increasing its probability of connection; and we introduce the grouping motif by ensuring that if a given expansion layer neuron connects to one member of a group the probability of connecting to others is higher. Again we use three input neurons, and pair two of them to create a group.

\begin{figure}[!ht]
    \centering
    \includegraphics[width=0.4\linewidth]{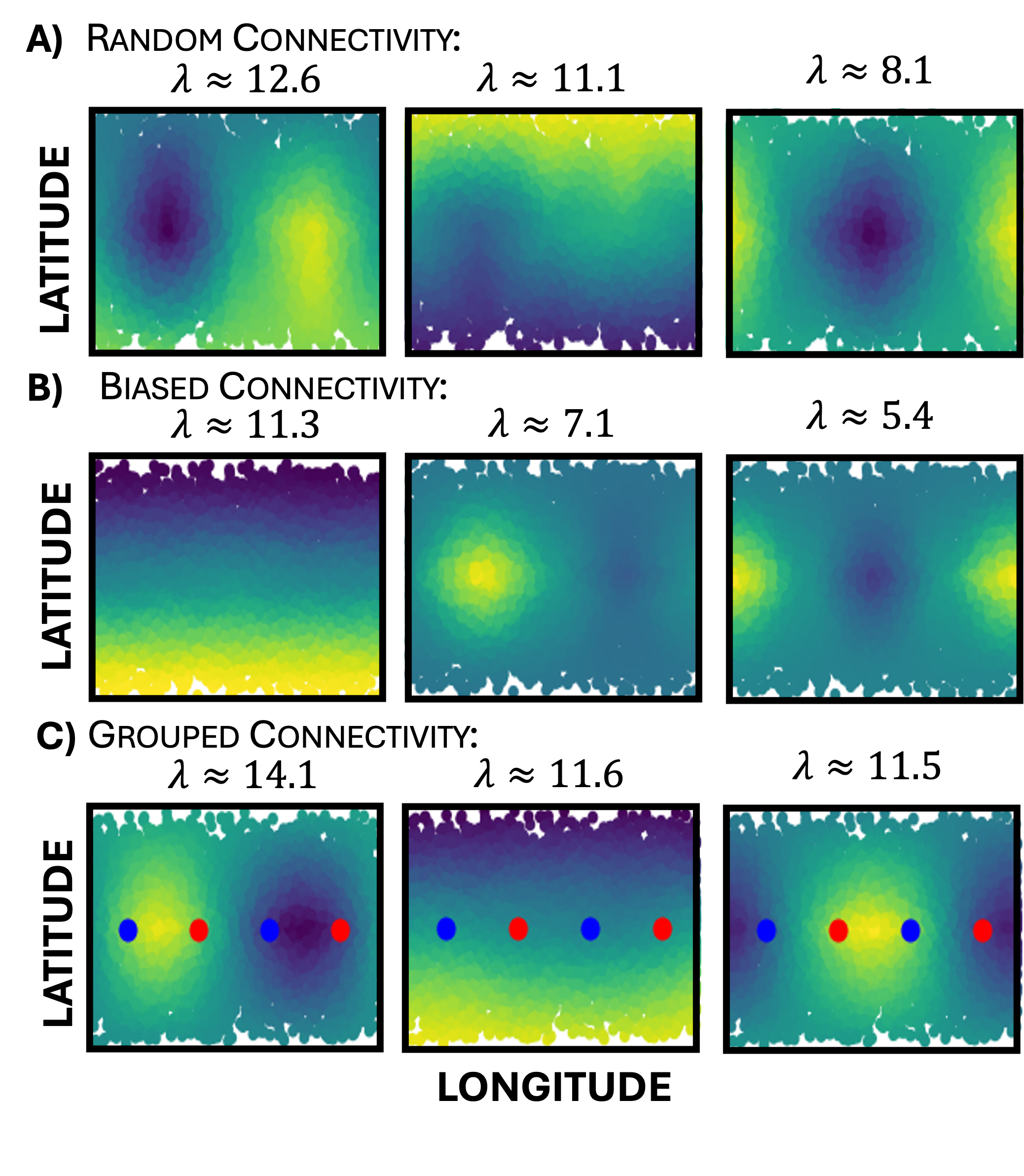}
    \caption{Rather than implementing the connectivity biases via correlations in weight values, we show that similar ideas generalise to a network where the biases in connectivity are implemented through connection probabilities, and with a more realistic nonlinearity. Here the connectivity matrix is sparse. The probability of a non-zero connection is either {\bf (A)} uniform, {\bf (B)} higher for one input neuron, or {\bf (C)} high for a particular input-hidden neuron pair if that hidden neuron is already connected to another input neuron in the same group -- in this case we again define the group to be the $x_1$ and $x_2$ input neurons. In each case we show the first three eigenvectors. {\bf A)} Networks are biased towards smooth functions, {\bf B)} but overconnecting in the $x_3$ direction creates biases towards variations along the $x_3$ direction. {\bf C)} Grouping the connectivity makes one mode in the grouped direction easier to learn, though interestingly the alignment of this mode changes relative to the previous section - red and blue points as in~\cref{fig:cerebellum-eigenfunctions}.}
    \label{fig:cerebellum_more_realisitic}
\end{figure}

In~\cref{fig:cerebellum_more_realisitic} we show that the same broad conclusions are true of this more plausible model as in the toy analytic case. In particular, randomly connected networks are biased towards smooth functions; networks overconnected to a given input find it easier to learn functions that rely on the activity of that input; and grouped connectivity leads to a bias towards functions that depend on the shared activity across members of the group. This network could be easily extended to explore other interesting effects. Simple plausibility fixes could include constraining the connectivity weights or the input activities to be positive, while more interesting neuroscience questions could include studying the impact of whitening pre-processing in the olfactory system \citep{wanner2020whitening}, or the impact of ongoing neurogenesis in the dentate gyrus \citep{seri2001astrocytes}.

\section{Discussion}

In this work we have studied the effect of non-random connectomic structure on the inductive bias of cerebellar-like networks. In particular, we have explored how a toy model of connectomic structure led to changes in the learning properties, via their effect on the kernel eigenstructure. In short, overconnection to some input neurons over others makes it easier for the circuit to learn functions that depend on that input, while groupings in the connectivity make it easy (hard) for the network to learn functions that generalise over (discriminate between) members of the same group \citep{zavitz2021connectivity,dorrell2023meta}. Relative to other works that study the same problem we provide mathematical tractability, linking the observed phenomena to well understood machine learning theory.

Broadly, this work aligns with approaches that seek to understand design choices in cerebellar-like networks through the tasks they help to solve \citep{harris2019additive,xie2022task,dorrell2023meta,zavitz2021connectivity}, in contrast to task-agnostic approaches that have sought to characterise cerebellar representations via measures such as the dimensionality of the representation \citep{litwin2017optimal,babadi2014sparseness,cayco2017sparse,cayco2019re}. In particular, understanding these learning effects via the kernel eigenfunctions has proven effective: beyond this work, increased representational sparsity has been linked to faster decay in the eigenvalues, producing systems that are tuned to learn smoother functions \citep{xie2022task}; and sparsity in the connectivity matrix has been shown to lead to networks that are inductively biased towards learning `low-order' functions -- those that depend on a small number of input dimensions \citep{harris2019additive,hashemi2023generalization}. Further, the same ideas have been used to understand visual representations \citep{bordelon2021population}. We find these account compelling. Excitingly, they are becoming experimentally testable. Genetic and pharmacological control now allows experimentalists to alter network parameters, such as the number of hidden layer neurons, connection density or sparsity \citep{ahmed2023hacking,elkahlah2020presynaptic}, and these experiments have confirmed predictions about the effect of network parameters on representational structure and behavioural discriminability. The predictions we make here could similarly be tested if a method to manipulate the connectivity covariance was found.


One shortcoming of understanding via a learner's inductive bias is the difficulty in interpretation for high-dimensional inputs. When the inputs are low-dimensional, simple pictures, like~\cref{fig:cerebellum-eigenfunctions}, suffice to understand a network's inductive bias; but in high-dimensions such pictures are hard to create and difficult to understand. It is in this setting in particular that analytic approaches hold promise: tractable toy models can be established that demonstrate the key phenomena while generalising to higher input dimensionalities. This can require contorting the problem into a solvable form while capturing the key biological phenomena, which is not always possible. That said, we are hopeful that recent developments in machine learning theory could usefully loosen the requirements for analytically tractability, providing dividends for understanding cerebellar-like networks \citep{hu2022universality}.

In sum, the theory of kernel regression provides a bridge from structure and function, allowing us to understand observed biological patterns through the corresponding learner's ability. We hope these approaches will generalise further.

\paragraph{Acknowledgements} The authors thank Pierre Glaser, Cengiz Pehlevan, Rodrigo Carrasco-Davis, Sina Tootoonian, the Schaeffer lab at the Francis Crick Institute, Maria Yuffa, Kevin Han Huang, Angus Silver, Dan Bush, and Basile Confavreux for conversations regarding this work, and the Gatsby Charitable Foundation for funding (GAT3755).

\printbibliography

\end{document}